| | |
|---|---|
| Title | Carbon nanotubes decorated with gold, platinum and rhodium clusters by injection of colloidal solutions into the post-discharge of an RF atmospheric plasma |
| Authors | N Claessens[1], F Demoisson[1,2], T Dufour[1], Ali Mansour[3], A Felten[4], J Guillot[3], J-J Pireaux[4] and F Reniers[1] |
| Affiliations | [1] Service de Chimie Analytique et Chimie des Interfaces (CHANI), Université Libre de Bruxelles, Faculté des Sciences, CP255, Boulevard du Triomphe 2, B-1050 Bruxelles, Belgium<br>[2] Laboratoire Interdisciplinaire Carnot de Bourgogne (ICB), D´epartement Nanosciences—EquipeMaNaPI: Matériaux Nanostructurés: Phénomènes à l'Interface,<br>UMR 5209 CNRS/Université de Bourgogne, 9 Avenue Alain Savary, BP 47870, F-21078 Dijon Cedex, France<br>[3] Department 'Science and Analysis of Materials' (SAM), Centre de Recherche Public—Gabriel Lippmann, 41 rue du Brill, L-4422 Belvaux, Luxembourg<br>[4] Facultés Universitaires Notre-Dame de la Paix, Centre de Recherche en Physique de la Matière et du Rayonnement (PMR), 61 rue de Bruxelles, B-5000 Namur, Belgium |
| Ref. | Nanotechnology, 2010, Vol. 21, Issue 38, 385603 |
| DOI | http://dx.doi.org/10.1088/0957-4484/21/38/385603 |
| Abstract | In this paper, we present a new, simple, robust and efficient technique to decorate multi-wall carbon nanotubes (MWCNT) with metal nanoparticles. As case studies, Au, Pt and Rh nanoparticles are grafted onto MWCNTs by spraying a colloidal solution into the post-discharge of an atmospheric argon or argon/oxygen RF plasma. The method that we introduce here is different from those usually described in the literature, since the treatment is operated at atmospheric pressure, allowing the realization in only one step of the surface activation and the deposition processes. We demonstrate experimentally that the addition of oxygen gas in the plasma increases significantly the amount of grafted metal nanoparticles. Moreover, TEM pictures clearly show that the grafted nanoparticles are well controlled in size. |

# 1. Introduction

Thanks to their unique electrical properties, high chemical stability and high surface-to-volume ratios, multi-wall carbon nanotubes (MWCNTs) have been proposed as ideal metal supports for catalytic and sensing applications. Therefore, many different methods have been developed to activate and decorate carbon nanotubes with metal nanoparticles. It is well accepted that the activation of a carbon surface is based on the formation of surface defects and/or the adsorption of reactive species, such as oxygen-containing groups. In most of the studies, this activation is carried out via wet chemistry techniques, requiring the use of hot nitric acid solutions [1]. In some others, the activation and deposition processes are based on vacuum techniques requiring ion guns, low pressure plasma and/or thermal evaporators [2].

In this work, we have successfully transposed a technique recently used for the deposition of Au nanoparticles onto HOPG (highly oriented pyrolytic graphite) [3] to the deposition of metal nanoparticles (Au [4], Pt and Rh) onto MWCNTs. Besides the advantage of working at atmospheric pressure, this technique guarantees the cleaning and the activation of the substrate surface in only one step [5]. Moreover, it allows the deposition of uniform, and well controlled in size, metal nanoparticles.

Another successful method that has been recently reported to decorate MWCNTs with gold, Rh and Pt nanoparticles is based on a thermal evaporation process [6–8]. Although the density of the grafted nanoparticles is higher than those we obtained, such a technique requires us to work at (ultra) high vacuum pressure. Moreover, the yield of this process is much lower than the one obtained when using the sprayed nanoparticles.

## 2. Materials & Experimental setup

MWCNTs were provided by Nanocyl S.A. (ref 3100). They were grown by chemical vapour deposition (CVD) with a purity higher than 95%. Their length was as high as 50 µm and their outer and inner diameters ranged respectively from 15 to 3 nm and 7 to 2 nm.

The metal nanoparticles to graft onto these MWCNTs stemmed from colloidal solutions. The gold colloidal solution was synthesized by the citrate thermal reduction method [3]. The colloidal solutions of platinum and rhodium were supplied by the Department of Chemistry (University of California) in Berkeley; their synthesis conditions are described in [9] and [10]. These suspensions resulted in stable dispersions of metal nanoparticles with an average diameter of 10 nm, 8 nm and 4 nm, respectively, for gold, rhodium and platinum.

Before starting any deposition, MWCNTs were soaked in a methanol solution for up to 15 min. Then, their surface was activated by exposure to the post-discharge of an atmospheric plasma during 2 min. This plasma was generated with an RF torch (Atomflo™-250 from Surfx Technologies LLC [11]) using argon as the vector gas. The decoration of the carbon nanotubes surface was performed by spraying the colloidal solution into the post-discharge (during 30 s). The resulting surface was then exposed to the post-discharge for 3 min. Once this step was achieved, the treated MWCNTs were introduced into a methanol solution for up to 5 min and subjected to ultrasonication. Their surface composition was finally analysed by x-ray photoelectron spectroscopy (XPS) and their surface morphology (dispersion and shape of the metal nanoclusters) was characterized by transmission electron microscopy (TEM).

In order to evaluate the chemical composition of the MWCNTs' surface, XPS analyses were performed by using two instruments. The HP 5950 A instrument from the LISE team (Facultés Universitaires Notre-Dame de la Paix, Namur, Belgium) was dedicated to the characterization of the Au/MWCNTs samples. The ThermoVG Microlab 350 instrument from the SAM team (CRPGL, Luxembourg) was used to analyze the Pt/MWCNT and Rh/MWCNT samples. In both cases, the RX source was a monochromatized Al Kα line (hν = 1486.6 eV). The samples were pressed onto a Cu tape on a holder and introduced into the spectrometer. The base pressure in the analytical chamber was ≈$10^{-9}$ mbar. Survey scans were used to determine the chemical elements present at the MWCNT surface. Narrow-region electron spectra were used for the chemical study of the C 1s, Au 4f, Pt 4f and Rh 3d peaks. The elemental composition was calculated after removal of a Shirley background line and using relative sensitivity factors from [12]. The obtained compositions must be taken as indicative and are used only for comparison between the different plasma treatments (with/without $O_2$) and the different deposited nanoparticles. They do not reflect the absolute surface composition. In our case, the multiply analyzed regions on the same sample always lead to the same relative compositions.

In order to determine the size and the distribution of the metal nanoclusters and to evaluate the effect of the plasma treatment on the MWCNTs, transmission electron microscopy (TEM) measurements were carried out on a Philips Tecnai 10 microscope operating at 80 kV from the LISE team for the characterization of the Au/MWCNT samples and on a LEO 922 OMEGA operating at 200 kV with a point-to-point resolution of 0.29 nm (TEM) from the SAM team for the characterization of the Pt/MWCNT and Rh/MWCNT samples. For this study, the MWCNT powder was dispersed in ethanol with ultrasonication for 5 min, and a drop was deposited onto a honeycomb carbon film supported by a copper grid. As a TEM image only gives local information, each sample was observed at different regions, so that each image presented in this paper is representative of the mean global observation.

## 3. Results & Discussion

As already mentioned in the experimental procedure, the decoration step was preceded by the exposure of the MWCNT surface to the post-discharge. During this plasma activation step, the sidewall structure of the carbon nanotubes can be modified. Figure 1 shows the XPS O 1s and C 1s core levels in three cases: no plasma activation, plasma activation (argon flow rate: 30 L.min$^{-1}$) and plasma activation using both argon and oxygen (with the respective flow rates: 30 L.min$^{-1}$ and 20 mL.min$^{-1}$). The carbon composition is 96.6% when no activation is performed. For the pure argon activation, this value decreases to 91.4% since the oxygen from the air was sufficient to create local oxygen vacancies. Then, for the Ar/O$_2$ plasma activation, the C 1s peak indicates a relative composition of only 88.1%, thus suggesting the significant impact of the oxygen during the plasma activation.

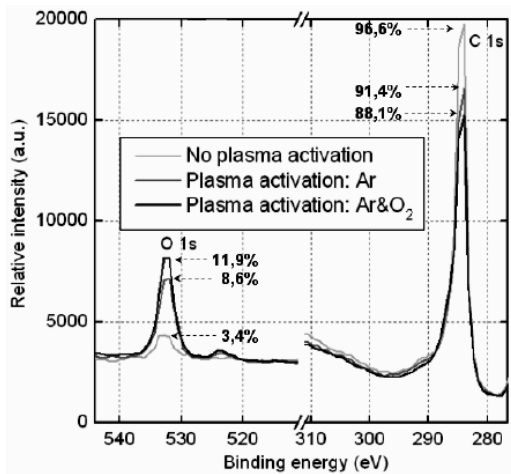

*Figure 1. XPS spectra core levels of O 1s and C 1s of MWCNTs: not activated, plasma-activated with an argon flow rate of 30 L.min$^{-1}$ and plasma-activated with an argon flow rate of 30 L.min$^{-1}$ and oxygen flow rate of 20 mL.min$^{-1}$. The RF plasma power was 80 W.*

TEM pictures were acquired to evaluate the impact of this Ar/O$_2$ plasma activation (figure 2(b)) with respect to the reference case where no plasma activation was performed (figure 2(a)). The comparison shows no significant damage on the external sidewalls that could be attributed to the plasma activation. In this sense, the HRTEM picture of the plasma-treated MWCNTs (figure 2(c)) shows their integrated layer structure, their smooth surface and the parallel graphene layers. Contrary to other studies where the carbon nanotubes are treated inside low pressure plasma, our MWNTs are exposed to the post-discharge of an atmospheric plasma. The ion-induced damage is thus strongly reduced [13].

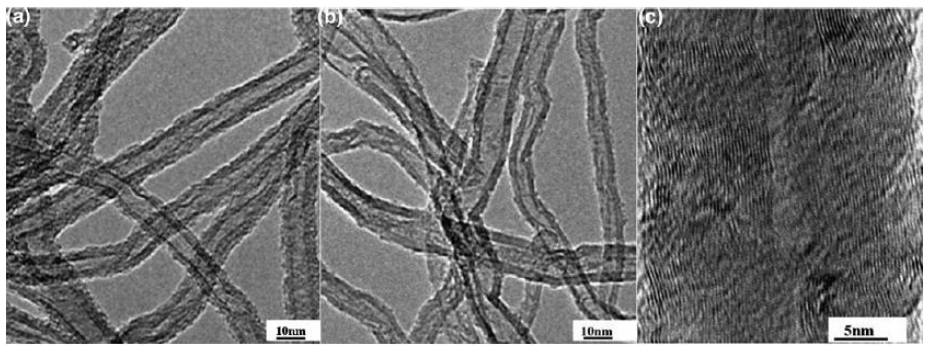

*Figure 2. TEM pictures of (a) pure MWCNTs and (b) MWCNTs plasma-activated with Ar and O$_2$. (c) is an HRTEM picture of case (b).*

After an argon plasma treatment with nanoparticles at atmospheric pressure, the XPS analysis highlighted the presence of metal on the MWCNTs' surface. The survey spectra (figure 3) indicated Au at. = 0.5%, Pt at. = 0.7% and Rh at. = 1.2%, respectively, for the Au/MWCNT, Pt/MWCNT and Rh/MWCNT samples. This surface composition was calculated with the usual formula, assuming therefore a homogeneous distribution of the elements in the sample (which is obviously not the case) and no matrix effects. With or without the ultrasonication step, the amount of metal detected was unchanged, thus suggesting a strong adhesion of the metal nanoparticles onto the MWCNTs. Note that the nitrogen detected on the carbon nanotubes decorated with Pt or Rh originates from the colloidal solution, and more specifically from its capping agent (polyvinylpyrrolidone). As according to the spectra of figure 3 the surface composition was approximately the same in any case, it is clear that gold, platinum and rhodium present a similarity of behavior This same surface composition also indicates a strong reproducibility of the plasma treatment for the metal cluster deposition.

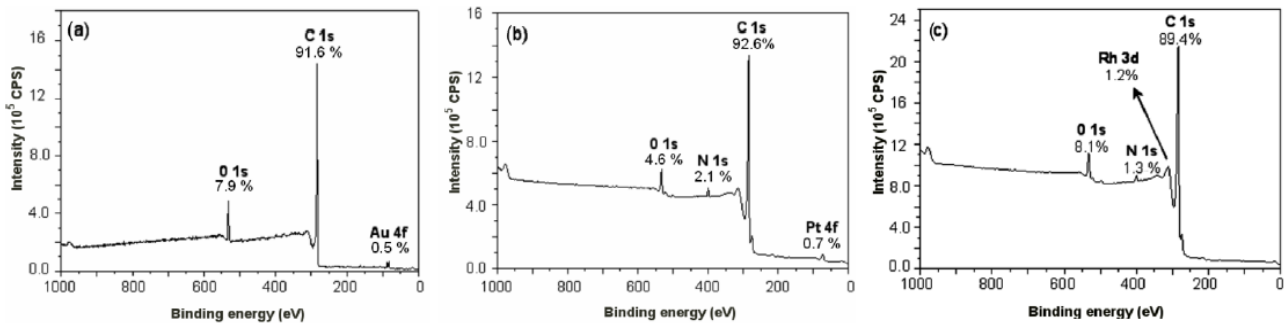

*Figure 3. Survey XPS spectra recorded on (a) Au/MWCNTs, (b) Pt/MWCNTs and (c) Rh/MWCNTs after atmospheric Ar plasma treatment.*

In figure 4, we show the fitted core levels of the Au 4f, Pt 4f, Rh 3d and C 1s peaks. The fitted Au 4f (Au 4f5/2–Au 4f7/2 spin–orbit doublets with a fixed 0.75:1 intensity ratio and 3.7 eV energy separation) [14], Pt 4f (Pt 4f5/2–Pt 4f7/2 spin–orbit doublets with a fixed 0.78:1 intensity ratio and 3.3 eV energy separation) [14] and Rh 3d (Rh 3d3/2–Rh 3d5/2 spin–orbit doublets with a fixed 0.64:1 intensity ratio and 4.75 eV energy separation) [14] spectra can be unambiguously assigned in the majority to the metallic state [15].

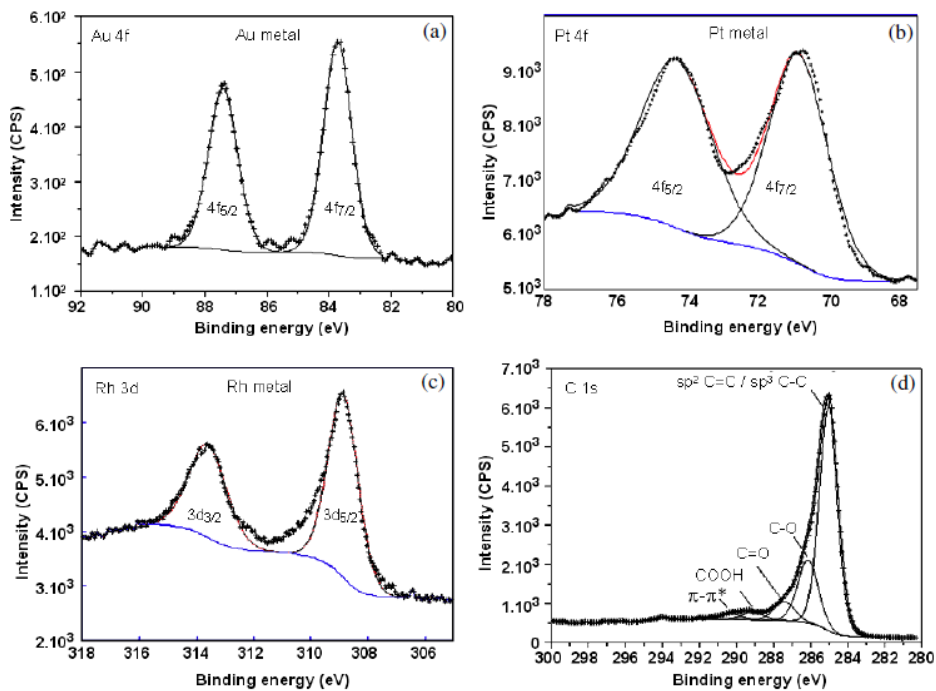

*Figure 4. Fitted core levels for (a) Au 4f, (b) Pt 4f, (c) Rh 3d and (d) C 1s peaks after the atmospheric argon plasma treatment. Note that the presented fitted C 1s level corresponds to Au/MWCNT sample characterization. The fitted C 1s level is similar for the Pt/MWCNT and Rh/MWCNT samples.*

As already observed on an HOPG surface [3], the high resolution C 1s spectra highlighted the presence of carbon-oxygen species. In the present case, illustrated in figure 4(d), the C 1s spectra can be fitted to five lines with binding energies at 284.9, 286.2, 287.4, 288.8 and 290.9 eV, with a constant FWHM of 1.3 eV [16]. These peaks correspond respectively to ($sp^2$ C=C)/($sp^3$ C–C), C–O, C=O, COOH and $\pi$–$\pi\ast$ contributions [4]. As the process does not operate in a closed system but in an open environment, the presence of the carbon–oxygen species can reasonably be attributed to the fact that the MWCNTs' surface was exposed to an oxidizing atmosphere during the plasma treatment. Indeed, optical emission spectrometry of the post-discharge (not shown) reveals the presence of atomic oxygen lines at 777 and 852 nm. Therefore, traces of oxygen and/or water cannot be avoided because of the humidity of the air. The solvent from the colloidal solutions (which is water) can also enhance the formation of these carbon–oxygen species.

The role played by these carbon–oxygen species is quite important on the amount of metal nanoparticles grafted onto the MWCNTs' surface. Theoretical calculations predicted that the surface must be in an appropriate energetic state to allow the condensation of metals on it. In the case of a decoration with gold nanoparticles, it has been calculated that the nucleation would preferably take place on defect sites, especially on oxygenated vacancies ($VO_2$) which would act as trapping sites for gold atoms [4].

To verify experimentally these predictions, we replaced the previous argon plasma treatment by a new one in which an oxygen flow rate of 20 ml min$^{-1}$ is added to the argon flow rate (30 L.min$^{-1}$). This new plasma treatment clearly modifies the surface composition of the MWCNTs, as illustrated by the XPS survey spectra in figure 5. The first spectrum (figure 5(a)) highlights about 4 at.% of gold deposited on the MWCNT surface instead of the 0.5 at.% detected without oxygen in the plasma gas (figure 3(a)). The potassium detected in the Au/MWCNT sample originates from the colloidal solution (potassium citrate was used in the synthesis process). In the case of platinum deposition (figure 5(b)), the absolute amount of grafted nanoparticles has considerably increased since the Pt 4f peak is much stronger than in figure 5(b). Finally, in figure 5(c) the relative surface composition of Rh is almost twice as high (2.2 at.% versus 1.2 at.%) when the post-discharge is enriched in oxygen. All these results are consistent and clearly show the benefit of mixing oxygen with argon, particularly in the case of gold deposition.

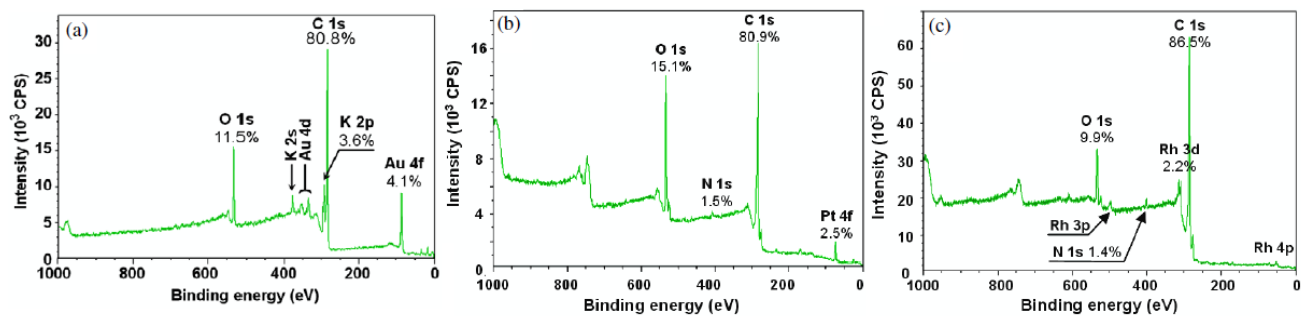

*Figure 5. Survey XPS spectra recorded on (a) Au/MWCNTs, (b) Pt/MWCNTs and (c) Rh/MWCNTs after the argon–oxygen plasma treatment. The flow rates were 30 L.min$^{-1}$ for argon and 20 mL.min$^{-1}$ for oxygen.*

In agreement with the previous XPS spectra (figure 5), we observed on the TEM images (figure 6) a higher number of gold nanoparticles grafted onto the MWCNT surface when oxygen is mixed with argon.

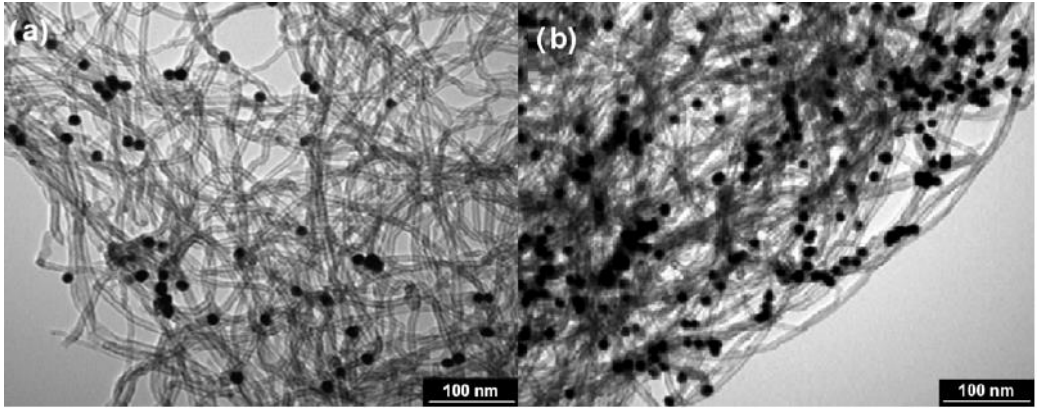

*Figure 6. TEM images of MWCNT samples covered with gold metallic nanoclusters (10 nm) deposited by the atmospheric plasma treatment using (a) Ar and (b) Ar/O$_2$ mixture. Flow rates were 30 L.min$^{-1}$ for Ar and 20 mL.min$^{-1}$ for O$_2$.*

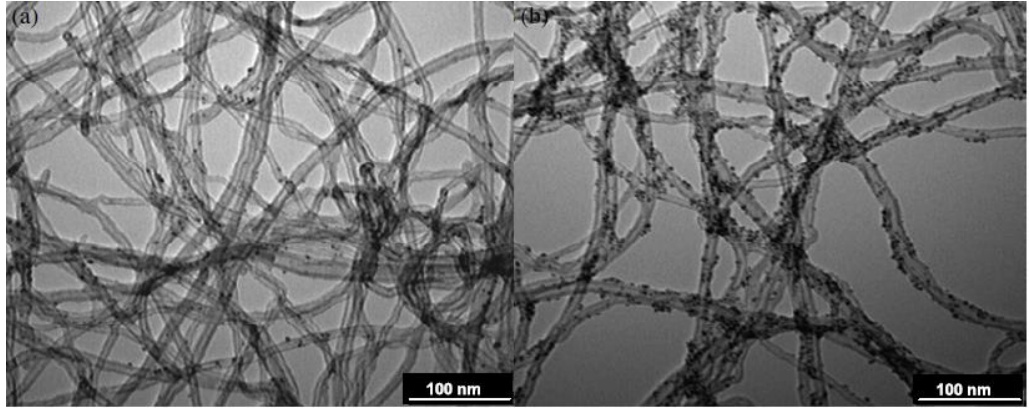

*Figure 7. TEM images of MWCNT samples covered with platinum metallic nanoclusters (4 nm in diameter) deposited by the atmospheric plasma treatment using (a) Ar and (b) Ar/O$_2$ mixture. Flow rates were 30 L.min$^{-1}$ for Ar and 20mL.min$^{-1}$ for O$_2$.*

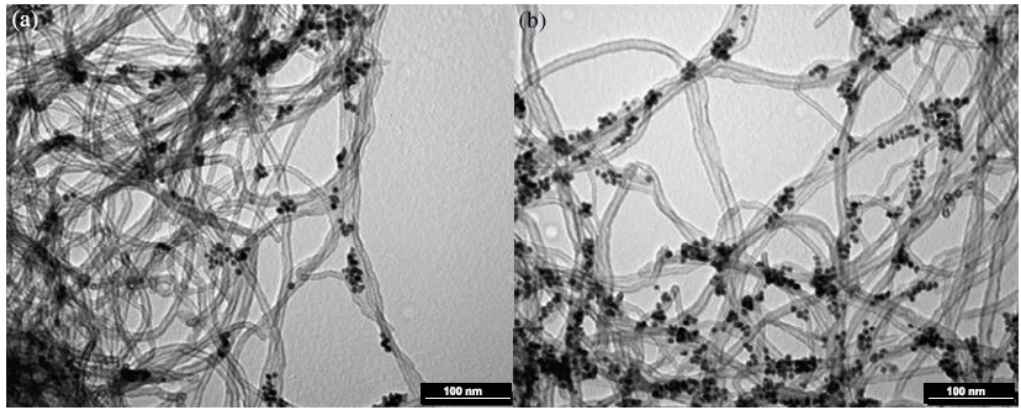

*Figure 8. TEM images of MWCNT samples covered with rhodium metallic nanoclusters (8 nm in diameter) deposited by the atmospheric plasma treatment using (a) Ar and (b) Ar/O$_2$ mixture. Flow rates were 30 L.min$^{-1}$ for Ar and 20mL.min$^{-1}$ for O$_2$.*

TEM images of MWCNTs decorated with platinum or rhodium nanoparticles are respectively presented in figures 7 and 8. In both cases, we obtained a similar increase in the number of metal nanoparticles grafted when operating under an argon–oxygen plasma treatment. This behaviour could be attributed to the increase of oxygen-containing groups and/or of structural defects on the MWCNTs' surface, under the assumption of a metal binding at a C–O or defect site [4]. TEM images (figures 6–8) show that most of the MWCNTs are isolated and that almost no agglomeration of carbon nanoparticles has occurred. Note that the diameters of spherical Au, Pt and Rh nanoclusters (respectively 10 nm, 4 nm and 8 nm) are similar to those of the preformed particles in the colloidal solution. Similar results have been obtained on HOPG substrates [3]. Finally, independent of their nature (Au, Pt or Rh), the dispersion of the nanoparticles was always better when an oxygen flow rate of 20 ml min−1 was mixed with the argon gas; and the grafted platinum nanoparticles were particularly homogeneously distributed (figure 7).

## 4. Conclusion

We have successfully achieved the deposition of gold, platinum and rhodium nanoparticles on MWCNTs by using an atmospheric plasma source. We have demonstrated the major role of oxygen during the plasma treatment. By adding an oxygen flow rate as low as 20 mL.min$^{-1}$ to the argon gas of the plasma torch, we managed to increase subsequently the amount of metal nanoparticles grafted onto MWCNTs (up to 4% in the case of gold), while preserving the original size of the nanoparticles from the colloidal solution. Our results point out a very interesting new and simple method which allows the deposition of metal nanoparticles with a desired uniform and well-controlled size.

## 5. Acknowledgments

The authors warmly thank Professor Gabor A Somorjai from the Department of Chemistry (University of California) in Berkeley, for supplying the colloidal solutions of rhodium and platinum. This work was financially supported under the European NMP3-CT-2006-033311 NANO2HYBRIDS project.

## 6. References

[1] Yu R, Chen L, Liu Q, Lin J, Tan K-L, Choon Ng S, Chan H S O, Xu G-Q and Andy Hor T S 1998 Chem. Mater. 10 718–22

[2] Ren G and Xing Y 2006 Nanotechnology 17 5596–601

[3] Demoisson F, Raes M, Terryn H, Guillot J, Migeon H-N and Reniers F 2008 Surf. Interface Anal. 566–70

[4] Charlier J-C et al 2009 Nanotechnology 20 375501

[5] Ladwig A, Babayan S, Smith M, Hester M, Highland W, Koch R and Hicks R 2007 Surf. Coat. Technol. 201 6460–4

[6] Felten A, Bittencourt C and Pireaux J J 2006 Nanotechnology 17 1954–9

[7] Bittencourt C, Hecq M, Felten A, Pireaux J J, Ghijsen J, Felicissimo M, Rudolf P, Drube W, Ke X and Van Tendeloo G 2008 Chem. Phys. Lett. 462 260

[8] Suarez-Martinez I, Ewels C P, van Tendeloo G, Thiess S, Drube W, Felten A, Pireaux J J, Ghijsen J and Bittencourt C 2010 ACS Nano at press

[9] Rioux R M, Song H, Hoefelmeyer J D, Yang P and Somorjai G A 2005 J. Phys. Chem. B 109 2192–202

[10] Wang Y, Ren J, Deng K, Gui L and Tang Y 2000 Chem. Mater. 12 1622–7

[11] Reniers F, Demoisson F and Pireaux J J 2009 Proc´ed´e de dépôt de nanoparticules sur un support PCT/EP2008/060676 patent

[12] Briggs D and Seah M P 1990 Practical Surface Analysis 2nd edn, vol 1 Auger and X-ray Photoelectron spectroscopy (New York: Wiley)

[13] Bittencourt C, Felten A, Douhard B, Colomer J-F, van Tandeloo G, DrubeW, Ghijsen J and Pireaux J J 2007 Surf. Sci. 601 2800–4

[14] NIST XPS Database http://srdata.nist.gov/xps/selEnergyType. aspx

[15] Buttner M and Oelhafen P 2006 Surf. Sci. 600 1170–7 http://techdb.podzone.net/xps-e/index.cgi?element=Pt, http://techdb.podzone.net/xps-e/index.cgi?element=Rh

[16] He X, Zhang F, Wang R and Liu W 2007 Carbon 45 2559–63